\documentclass[aps,prl,twocolumn,unsortedaddress,floats,showpacs]{revtex4}
\usepackage{graphicx}
\usepackage{amssymb}
\usepackage{amsmath}
\usepackage{multirow}
\usepackage{color}

\begin{document}

\title{Charging and heat collection by a positively charged dust grain in a plasma}

\author{Gian Luca Delzanno}
\email{delzanno@lanl.gov}
\author{Xian-Zhu Tang}
\email{xtang@lanl.gov}
\affiliation{Theoretical Division,
Los Alamos National Laboratory,
Los Alamos, New Mexico 87545}

\date{\today}

\begin{abstract}
  Dust particulates immersed in a quasineutral plasma can emit
  electrons in several important
  applications. Once electron emission becomes strong enough, the
  dust enters the positively charged regime where the conventional
  Orbital-Motion-Limited (OML) theory can break down due to
  potential well effects on trapped electrons.  A minimal modification
  of the trapped-passing boundary approximation in the so-called
  OML$^+$ approach is shown to accurately predict the dust charge and heat
  collection flux for a wide range of dust size and temperature.
\end{abstract}

\maketitle

The problem of the charging of a solid body in a plasma has a long
history and various applications ranging from probes and
spacecraft to planet formation to dusty plasmas in laboratory and
space \cite{shukla01}. The body collects plasma particles and is often negatively charged owing to
the higher electron mobility. In many instances, however, the body can
emit electrons (via thermionic emission,
photoemission and secondary electron emission) and become positively charged. Examples of such
include spacecraft applications~\cite{whipple81}; the moon~\cite{poppe11}; 
meteoroids entering the Earth's atmosphere~\cite{sorasio01}; ionospheric rockets experiments~\cite{macdonald06}; 
dust particles in the solar wind~\cite{kimura98}, planetary rings~\cite{horanyi96}, cometary
environments~\cite{klumov07}, or magnetic fusion
devices~\cite{smirnov07,vaverka}. Experiments showing the formation of ordered
structures with positively charged dust are reported in laboratory~\cite{fortov96,samarian01} and in microgravity~\cite{fortov98}. 
Since charging is governed by the characteristic length of the body
relative to the plasma { Debye length or the electron gyroradius}, there is no conceptual difference between the examples above and 
in what follows we will use the term dust broadly.

A charging theory is a necessary ingredient of any
model of dust transport and
destruction/survival in a plasma \cite{mendis94,sorasio01,kimura98,horanyi96,ticos06,smirnov07,tang10,bacharis10,ratynskaia13,delzanno13}.
It calculates the dust charge/potential, momentum and heat collection due to the dust-plasma interaction. The most widely used charging theory is the
Orbital-Motion-Limited (OML) theory \cite{mottsmith26}, which leads to a simple nonlinear equation for the dust
potential.  

In this Letter, we use PIC simulations and theoretical
analysis to show that OML can become inapplicable in the positively
charged regime.  It can completely miss the transition between negatively and
positively charged dust (thus predicting a positive dust potential
when simulations show a negative dust potential) and overestimates
the power collected by the grain (up to a factor of $2$ for the
cases considered). This is due to the development of a non-monotonic
potential (a potential well) near the grain. The fact that a potential
well can exist near an electron emitting body is known~\cite{hilgers04,delzanno04,poppe11}. However, this is the first study that illustrates
the breakdown of OML in this regime. Moreover, this Letter presents a revised
charging theory which is as simple as OML, recovers OML in the
appropriate limits, but remains accurate when potential well
effects are important.

We study the charging of a spherical dust grain of radius $r_d$ at rest in
a collisionless, unmagnetized hydrogen plasma ($m_{e\,(i)}$ and $T_{e\,(i)}$ are the electron (ion) mass
and temperature, $n_{0}$ is the unperturbed plasma density away from the grain). The dust grain charges by collecting
plasma and emitting electrons. The dynamics of the system is governed by the electric field created by the charged dust and 
a dynamical equilibrium is reached where the sum of all the currents on the dust surface is zero (floating condition).

In order to understand the OML limitations, we recall that the steady
state of the system under consideration is completely determined by the {\it orbital motion} (OM)
theory~\cite{laframboise66,kennedy03,delzanno05}. OM is based on the
conservation of energy and angular momentum
\begin{align}
  v_r^2+v_\theta^2-\frac{2 e}{m_e} \phi_d &
  =v_r^{\prime2}+v_\theta^{\prime2}-\frac{2 e}{m_e} \phi(r),
\label{E1} \\
m_e r_d v_\theta & =m_e r v_\theta^\prime,
\label{L1}
\end{align}
and the conservation of the number of particles along
characteristics in phase space (Liouville's theorem).  Here
we have introduced a spherical reference frame centered on the dust
grain where $r$ is the radial distance and $v_r$ ($v_\theta$) is the radial (tangential) velocity of a particle, 
$e$ is the elementary charge, $\phi$ is the electrostatic potential, and
$\phi_d=\phi(r_d)$.  Equations (\ref{E1}) and (\ref{L1}) are
for the emitted electrons, but similar relations hold for the
background plasma.  They can be
combined into
$v_r^2-U(r,v_\theta)=v_r^{\prime2}$,
stating the conservation of energy for a particle moving radially
in the effective potential
$U(r,v_\theta)\equiv \frac{2 e}{m_e} \left[\phi_d - \phi(r) \right]-\left[1-\left(\frac{r_d}{r} \right)^2 \right] v_\theta^2=F_{E}-F_{C}$.
The first (second) term of $U(r,v_\theta)$ is due to the electrostatic (centrifugal) force $F_{E}$ ($F_{C}$).

When electron emission is not dominant, the dust grain is negatively
charged, $\phi(r)$ is monotonic, and all the emitted
electrons leave the grain (labeled as passing electrons) and contribute to the emitted current.  
In this regime OML approximates OM by
neglecting potential barriers to the ion motion associated with
maxima of the ion effective potential~\cite{alpert65,allen00}.

When electron emission is dominant, the dust grain is positively
charged and $\phi(r)$ is non-monotonic: the slowest
emitted electrons are attracted back to the grain creating a trapped
electron population~\cite{delzanno04,delzanno05}. { Here 'trapped' refers to
those emitted electrons that are re-collected by the grain, and not to particles
on a confined/bounded orbit as in probe theory~\cite{laframboise66,goree92,allen00}}. The emitted
electrons experience potential barriers to their motion: depending on
$v_\theta$, the effective potential can have a maximum or be
monotonically decreasing. The position of the maximum $r_m$ is given by
$-\frac{e}{m_e} r_m^3 \phi^\prime(r_m)=r_d^2 v_\theta^2$, which has
one solution only for $r_d\le r_m \le r_{min}$ [the
minimum of $\phi(r)$ is $\phi_{min}=\phi(r_{min})$], namely when $\phi^\prime<0$. For $v_\theta=0$, the maximum
is at $r_m=r_{min}$.  One can therefore define a critical tangential
velocity, $v_\theta^*=\sqrt{-\frac{e}{m_e} r_d \phi^\prime_d}$
\cite{delzanno05}, to characterize the electron orbits around
the grain. For $v_\theta>v_\theta^*$ the effective potential is
monotonically decreasing ($F_C\gg F_{E}$): all the emitted electrons 
leave the grain, irrespective of their radial velocity. For $v_\theta<v_\theta^*$, the effective
potential has a maximum ($F_E\gtrsim F_{C}$): only those electrons with radial
velocity $v_r>\sqrt{U(r_m(v_\theta),v_\theta)}$ leave the grain and
contribute to the net current.  Thus, the OM
trapped/passing boundary (TPB) for evaluating the
dust electron emission current is
\begin{equation}
v_r^2=U(r_m(v_\theta),v_\theta).
\label{tpbom}
\end{equation}
When the emitted electrons follow a
Maxwellian distribution (representative of most applications) with
temperature $T_d$ and thermal speed $v_{th,t}=\sqrt{T_d/m_e}$, the implication of Eq.~(\ref{tpbom}) can be elucidated for
$ v_\theta^*/v_{th,t} \gg 1$. Most of
the emitted electrons have velocity $v\lesssim v_{th,t}$ and see a
potential barrier located at $r_m\simeq r_{min}$ corresponding to the
TPB given by $v_r^2=U(r_{min},v_\theta)$, i.e.
\begin{equation}
  v_r^2+\left[1-\left(\frac{r_d}{r_{min}}\right)^2\right]v_\theta^2=\frac{2 e}{m_e} \left(\phi_d - \phi_{min} \right).
\label{tpbappr}
\end{equation}
This produces an ellipsoid in velocity space with aspect ratio
given by $1-\left(\frac{r_d}{r_{min}}\right)^2.$ 

\begin{figure}
\centering
\includegraphics[scale=0.43]{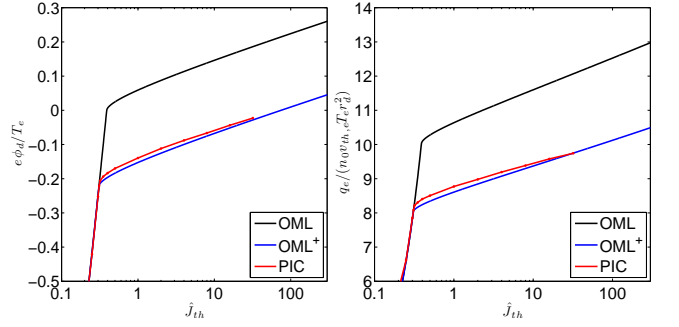}
\caption{Dust potential (left) and power collected by the dust from the background electrons (right)
versus emitted current [$T_d/T_e=0.03$, $r_d/\lambda_{De}=1$].}
\label{fig1}
\end{figure}

Strictly speaking, OML always assumes a monotonic $\phi(r),$ so the trapping of emitted electrons is only possible
if $\phi_d > 0.$ Adopting Sodha's formula for positively charged dust
in vacuum~\cite{sodha71}, the OML TPB is
\begin{equation}
v_r^2+v_\theta^2=\frac{2 e}{m_e} \phi_d.
\label{tpboml}
\end{equation}
Contrasting Eqs. (\ref{tpbappr}) and (\ref{tpboml}), it is clear that
the OML TPB approximation is only resonable for $|\phi_d| \gg |\phi_{min}|$
and $r_d \ll r_{min}$.
The discrepancy can lead to drastically different predictions for $\phi_d$ and
the power $q_e$ collected by the dust grain from the background electrons,
see Fig.~\ref{fig1}.

The OM result in Fig.~\ref{fig1} is obtained by a self-consistent electrostatic
PIC simulation.  This is for a perfectly conducting
spherical dust grain as the inner boundary of the simulation domain,
while the outer boundary is a concentric sphere of radius $R$.  The
dust grain emits electrons by thermionic emission, which are modeled
with a Maxwellian distribution, $f_{th}({\bf v})=2
\frac{m_e^3}{h^3} \exp \left(- \frac{m_e {\bf v}^2}{2 T_d} -
  \frac{W}{T_d}\right),$ with $h$ Planck's constant and $W$ the dust
thermionic work function.  For a negatively charged dust grain, 
the thermionic current density (normalized to $e n_0 v_{th,e}$ with $v_{th,e}=\sqrt{T_e/m_e}$) is given by the
Richardson-Dushman formula \cite{ashcroft}
\begin{equation}
\hat{J}_{th}=\frac{J_{th}}{e n_0 v_{th,e}}=\frac{4 \pi m_e T_d^2}{n_0 v_{th,e} h^3} \exp\left(-\frac{W}{T_d}\right).
\label{dushman}
\end{equation}
The smallest time step of the simulations is $\Delta t \omega_{pe}=0.0125$ ($\omega_{pe}=\sqrt{e^2 n_0 /(\varepsilon_0 m_e)}$ with 
$\varepsilon_0$ vacuum permittivity), while other parameters are
$R/\lambda_{De}=10$ ($\lambda_{De}=\sqrt{\varepsilon_0 T_e/(n_0 e^2)}$), $m_i/m_e=1836$, and $T_i/T_e=1$. More details on the simulation model can be found
in Ref.~\cite{delzanno04}.

\begin{figure}
\centering
\includegraphics[scale=0.35]{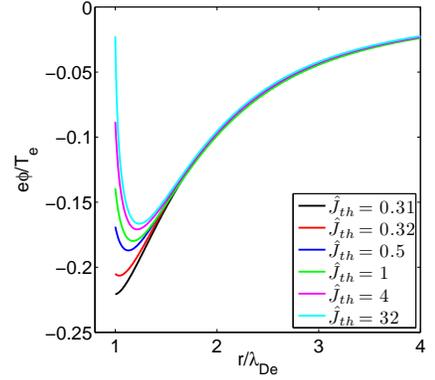}
\caption{Potential for various $\hat{J}_{th}$ ($T_d/T_e=0.03$, $r_d/\lambda_{De}=1$). }
\label{fig2}
\end{figure}

The OML result in Fig.~\ref{fig1} is obtained using
Eq.~(\ref{dushman}) when $\phi_d<0$, and by taking into account the
trapping of emitted electrons when $\phi_d>0$ according to Eq.~(\ref{tpboml}) \cite{sodha71}.  The PIC and OML
predictions are contrasted in Fig.~\ref{fig1} for $r_d/\lambda_{De}=1$
and $T_d/T_e=0.03$. For reference, without thermionic emission the
grain is negatively charged with good agreement between theory and
simulations: $e\phi_d^{{\rm OML}}/T_e\simeq -2.50$ while
$e\phi_d^{{\rm PIC}}/T_e\simeq -2.54$ \cite{delzanno13ieee}. The
curves in Fig.~\ref{fig1} (left) exhibit the characteristic behavior
associated with increasing thermionic emission: initially there is a
sharp increase of the dust potential since the grain is becoming less
negatively charged. All the emitted electrons leave the grain (hence
the agreement with OML).  At a critical current the curves bend since
the dust grain is now positively charged and a population of trapped
emitted electrons exists.  PIC simulations show that the transition
from negatively to positively charged grain occurs at negative dust
potential [$e{\phi_d^*}^{{\rm PIC}}(\hat{J}_{th}^*\simeq
0.3)/T_e\simeq -0.22$].  Figure~\ref{fig1} (right) shows the
(normalized) power collected by the dust from the background electrons
$\hat{q}_e=q_e/\left(n_0 v_{th,e} T_d r_d^2\right)$. This is a
particularly important quantity for dust survivability in a plasma,
since positively charged grains are heated almost exclusively by the
background electrons.  As the negatively charged particle starts to
emit thermionically, it reduces its charge and repels less background
electrons. These electrons heat the dust particle, which emits more
and further lowers its charge, and a positive feedback is established
that can lead to dust destruction \cite{smirnov07}. One can see in
Fig.~\ref{fig1} that OML overestimates $\hat{q}_e$ by $\sim30\%$.
{ Finally, the conventional relation between dust charge $Q_d$ and
potential, $Q_d=4 \pi \varepsilon_0 r_d \phi_d$, no longer holds (details will be
presented elsewhere).}  As expected, these
discrepancies are due to a deep and localized potential well
(Fig.~\ref{fig2}).

\begin{figure}
\centering
\includegraphics[scale=0.35]{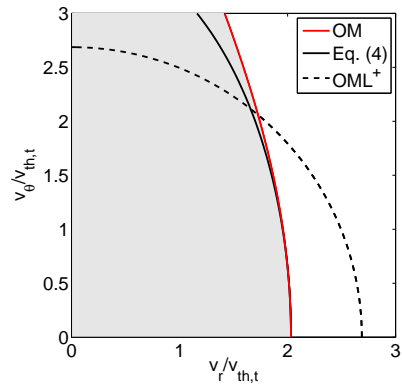}
\caption{Thermionic electrons trapped/passing boundary for 
$T_d/T_e=0.03$, $\hat{J}_{th}=2$ and $r_d/\lambda_{De}=1$.}
\label{om}
\end{figure}

The equilibrium potential obtained by PIC can be used to
check the emitted electrons TPB for OML and OM as shown in Fig. \ref{om} for $\hat{J}_{th}=2$, $r_d/\lambda_{De}=1$
and $T_d/T_e=0.03.$ For OM, $r_m(v_\theta)$ is calculated numerically
and the TPB is given by Eq.~(\ref{tpbom}). 
The approximate form, Eq.~(\ref{tpbappr}), is reasonably close in this range.
For $\hat{J}_{th}=2$ one has $\phi_d^{\rm PIC}<0$ so the OML approximation would have had
passing electrons for the entire $(v_r,v_\theta)$ space.  This cannot
yield a solution since the OML currents cannot balance for the correct $\phi_d^{\rm PIC}.$ Instead, OML
forces an incorrect prediction of a positive $\phi_d^{OML}$ in order
to reduce (relative to $\hat{J}_{th}$) the emitted electron current. 

\begin{figure}
\centering
\includegraphics[scale=0.65]{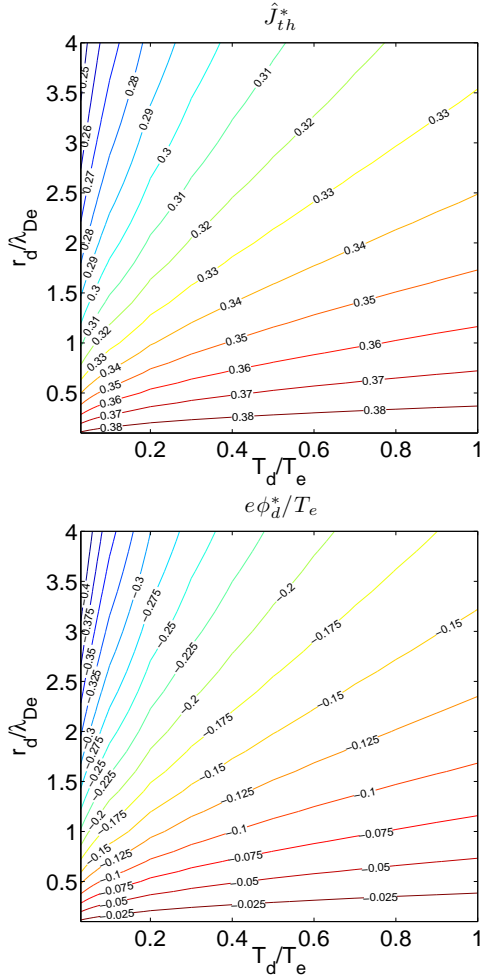}
\caption{Contours of $\hat{J}_{th}^*$ and $e\phi_d^*/T_e$ varying $T_d/T_e$ and $r_d/\lambda_{De}$.}
\label{scan}
\end{figure}

Can we anticipate when potential well effects are important and OML becomes inaccurate?
This is set by the critical thermionic current $\hat{J}_{th}^*$
corresponding to $Q_d=0$, below which OML is still accurate when $r_d\sim \lambda_{De}$ \cite{willis10}. 
We find the solution by solving Poisson's equation
$\nabla^2 \phi=\frac{e}{\varepsilon_0} \left[n_e^{OML}(\phi)-n_i^{OML}(\phi)+n_{th}(\phi)\right]$.
The OML electron density can be found in Refs.~\cite{alpert65,allen00} while a new expression for the ion density will be presented elsewhere \cite{tang14}.  The density of the (passing)
thermionic electrons is 
\begin{eqnarray}
  &&\frac{n_{th}}{n_0}=\sqrt{\frac{\pi}{2}}\frac{\hat{J}_{th}}{\sqrt{\frac{T_d}{T_e}}} \exp \left(e\frac{\phi-\phi_d}{T_d}\right)
  \left[ 1- {\rm erf} \sqrt{e \frac{\phi-\phi_d}{T_d}}\right. - \nonumber \\
  && \left.\exp\left[\frac{e\left(\phi-\phi_d\right)}{T_d\left(z^2-1\right)}\right]\frac{\sqrt{z^2-1}}{z}
    \left(1-{\rm erf}\sqrt{\frac{e\left(\phi-\phi_d\right) z^2}{T_d\left(z^2-1\right)}} \right)\right], \nonumber
\end{eqnarray}
where $z=r/r_d$.
Poisson's equation is then solved with conditions
$\phi(r_d)=\phi_d^{OML}$ and $\phi^\prime(R\gg r_d)=0$, while the additional
constraint $Q_d\propto\phi^\prime|_{r_d}=0$ is used to determine
$\hat{J}_{th}^*$. For $T_d/T_e=0.03$ and $r_d/\lambda_{De}=1$, we
obtain $\hat{J}_{th}^*=0.31$, in excellent agreement with
Fig. \ref{fig2}.  For their practical importance, we plot the contours
of $\hat{J}_{th}^*$ and $e\phi_d^*/T_e$ varying
$T_d/T_e$ and $r_d/\lambda_{De}$ in Fig. \ref{scan}.  The value of
$\phi_d^*$ is representative of the importance of potential well effects: 
the higher $|e\phi_d^*/T_e|$, the more important
these effects are. Figure~\ref{scan} shows that for $r_d \ll \lambda_{De}$ potential well effects
are unimportant and OML is also a good approximation when $Q_d>0$. On the other hand, as $r_d$
increases these effects become important, more so if $T_d/T_e$ is
small since the potential well is deeper and more localized. Conditions where potential well effects could be important
are easily met in magnetic fusion applications~\cite{delzanno13} and can
be met for mm-sized (and above) meteoroids entering the Earth's atmosphere~\cite{sorasio01}.

Is it possible to develop an accurate approximation to OM as simple as OML when potential well effects are
important?  This is a challenge since Poisson's equation is not solved for $\phi(r)$
in an OML-like approach, so the promising approximation~(\ref{tpbappr}) cannot be deployed
for lack of information on $r_{min}$ and $\phi_{min}.$
Extending Sodha's idea~\cite{sodha71}, an OML-like TPB approximation
would have $\phi_d$ in Eq.~(\ref{tpboml}) replaced by $\phi_d - \phi_d^*$,
since for $\phi_d>\phi_d^*$ the dust becomes positively charged and starts
trapping some of the emitted electrons.
So the revised OML-like TPB is
\begin{align}
v_r^2+v_\theta^2=\frac{2 e}{m_e} \left(\phi_d -\phi_d^*\right).
\label{tpbomlp}
\end{align}
From Fig.~\ref{om}, one can see that Eq.~(\ref{tpbomlp}) approximates
the OM TPB in a least squares sense.  From
Eq.~(\ref{tpbomlp}), the net thermionic current becomes:
\begin{equation}
I_{th}= 4 \pi r_d^2 J_{th} \exp \left[-\frac{e\left(\phi_d-\phi_{d}^*\right)}{T_d} \right] \left[1+\frac{e\left(\phi_d-\phi_{d}^*\right)}{T_d} \right].
\label{Ithtot}
\end{equation}
Thus, for $Q_d>0$ ($\hat{J}_{th}>\hat{J}_{th}^*$),
Eq.~(\ref{Ithtot}) will be used to calculate the dust potential from
the floating condition.  The electron and ion background collection currents remain those of OML.  We refer to approximation~(\ref{Ithtot}) as OML$^+$, since it modifies OML only for $Q_d>0$.

The results from OML$^+$ are plotted in Fig.~\ref{fig1},
where one can see that the agreement with
PIC simulations is very good.  For instance, for
$\hat{J}_{th}=32$, OML$^+$ gives $e\phi_d/T_e\simeq-0.028$, while
$e\phi_d^{PIC}/T_e\simeq -0.023$.  For comparison, 
$e\phi_d^{OML}/T_e\simeq 0.19$. For higher $T_d$, for instance $T_d/T_e=0.5$,
the potential well effect is not very important and both OML and OML$^+$
capture the dust potential reasonably (the
relative error is less than $15\%$, not shown).  We have also investigated the
effect of $r_d$ for $\hat{J}_{th}=2$
and $T_d/T_e=0.03$, as shown in Table~\ref{t2}.  As
expected, the dust potential is independent of $r_d$ in
OML. In OML$^+$, however, the $r_d$ dependence is taken
into account through $\phi_d^*$, resulting in a remarkable
agreement with PIC: in the cases studied the relative
error between OML$^+$ and PIC is  $\sim10\%$. Furthermore, the
discrepancy between OML and PIC widens as $r_d$ increases, signaling that potential well effects are becoming
more important [cf. Fig.~\ref{scan}].  Table \ref{t2} also shows $\hat{q}_e$: for $r_d/\lambda_{De}=4$,  $\hat{q}_e^{OML^+}\simeq \hat{q}_e^{PIC} \simeq7.4$,
while $\hat{q}_e^{OML}\simeq 11$.  Hence, in this case OML
overestimates the dust collected power by roughly
$50\%$. In magnetic fusion energy applications such discrepancy can mean predicting dust
destruction instead of survival, with strong implications for the
safety and performance of the machine~\cite{delzanno13}.

\begin{table}
\centering
\caption{Parametric study increasing $r_d$ for $\hat{J}_{th}=2$ and $T_d/T_e=0.03$ ($R/\lambda_{De}=20$ for $r_d/\lambda_{De}>1$).}
\begin{tabular}{|c|c|c|c|c|c|c|c|c|}
\hline
 $r_{d}/\lambda_{De}$ & $e\phi_d^{PIC}/T_e$ & $\hat{q}_e^{PIC}$& $\hat{J}_{th}^*$ & $e\phi_d^*/T_e$ & $e\phi_d^{{\rm OML}^+}/T_e$ & $\hat{q}_e^{{\rm OML}^+}$\\ \hline
$1$ & $-0.113$ & $9.0$&$0.31$ & $-0.22$ & $-0.125$ & $8.8$\\
$2$ & $-0.208$ & $ 8.1$&$0.27$ & $-0.33$ & $-0.228$ & $8.0$\\
$3$ &  $-0.268$& $7.7$& $0.26$ & $-0.39$ & $-0.292$ & $7.5$\\
$4$ & $-0.309$ & $7.4$&$0.24$ & $-0.44$ & $-0.336$ & $7.2$\\
\hline
\end{tabular}
\label{t2}
\end{table}

In conclusion, OML theory can break down in the positively charged
regime for electron-emitting dust because it does not 
account for potential well effects on the TPB of the
emitted electrons.  Remarkably, once the critical dust potential
$\phi_d^*$ at which the dust becomes positively charged is found by OML, a
revised trapped-passing boundary for electron emission,
Eq.~(\ref{tpbomlp}), yields an OML$^+$ approximation that accurately
predicts $\phi_d$ and dust power collection $q_e$ over a wide range of
dust size $r_d/\lambda_{De}$ and temperature $T_d/T_e$.  With
$\phi_d^*$ from Fig.~\ref{scan}, OML$^{+}$ can be
readily deployed by substituting Eq.~(\ref{Ithtot}) into
the conventional OML theory.

\medskip{}
This work was funded by the Laboratory Directed Research and Development (LDRD) program,
U.S. Department of Energy Office of
Science, Office of Fusion Energy Sciences, under the auspices of the
National Nuclear Security Administration of the U.S. Department of
Energy by Los Alamos National Laboratory, operated by Los Alamos
National Security LLC under contract DE-AC52-06NA25396.

\bibliography{biblio2}

\end{document}